\documentclass[12pt]{article}

\usepackage{amsmath,amsfonts,amsthm,verbatim,hyperref,graphicx}

\usepackage{epstopdf}
\usepackage{centernot}
\usepackage{authblk}
\usepackage{caption}
\newcommand{\A}{\mathrm{A}}
\newcommand{\B}{\mathrm{B}}

\DeclareMathOperator{\Prob}{\mathbb{P}}
\DeclareMathOperator{\E}{\mathbb{E}}
\DeclareMathOperator{\Harmonic}{H}

\newcommand{\klow}{\bar{\kappa}_\mathrm{low}}
\newcommand{\khigh}{\bar{\kappa}_\mathrm{high}}
\newcommand{\kA}{{\kappa}_\mathrm{A}}
\newcommand{\kH}{{\kappa}_\mathrm{H}}

\newcommand{\vx}{\mathbf{x}}

\newtheorem*{theorem}{Theorem}

\DeclareMathSymbol{\cart}{\mathbin}{AMSa}{"03}

\title{Evolutionary games on isothermal graphs}

\author[a,b]{Benjamin Allen\thanks{These authors contributed equally to this work}}
\author[c]{Gabor Lippner$^*$} 
\author[b,d,e]{Martin A.~Nowak} 

\affil[a]{Department of Mathematics, Emmanuel College, Boston, MA, USA}
\affil[b]{Program for Evolutionary Dynamics, Harvard University, Cambridge, MA, USA}
\affil[c]{Department of Mathematics, Northeastern University, Boston, MA, USA}
\affil[d]{Department of Mathematics, Harvard University, Cambridge, MA, USA}
\affil[e]{Department of Organismic and Evolutionary Biology, Harvard University, Cambridge, MA, USA}

\begin{document}
\maketitle

\begin{abstract}
Population structure affects the outcome of natural selection. Static population structures can be described by graphs, where individuals occupy the nodes, and interactions occur along the edges. General conditions for evolutionary success on any weighted graph were recently derived, for weak selection, in terms of coalescence times of random walks. Here we show that for a special class of graphs, the conditions for success take a particularly simple form, in which all effects of graph structure are described by the graph's ``effective degree"---a measure of the effective number of neighbors per individual. This result holds for all weighted graphs that are isothermal, meaning that the sum of edge weights is the same at each node.  Isothermal graphs encompass a wide variety of underlying topologies, and arise naturally from supposing that each individual devotes the same amount of time to interaction. Cooperative behavior is favored on a large isothermal graph if the benefit-to-cost ratio exceeds the effective degree. We relate the effective degree of a graph to its spectral gap, thereby providing a link between evolutionary dynamics and the theory of expander graphs. As a surprising example, we report graphs of infinite average degree that are nonetheless highly conducive for promoting cooperation.
\end{abstract}

The structure of a population has important consequences for its evolution \cite{wright1943isolation,kimura1964stepping,comins1992spatial,HanskiGilpin,TilmanSpatial,ErezGraphs,NowakStructured,allen2015molecular}.  In particular, spatial or social network structure can promote the evolution of cooperative behavior, by allowing cooperators to cluster together and share benefits \cite{NowakMay,DurrettLevin,LionvanBaalen}.

Spatial structure can be represented mathematically as a graph or network, in which nodes represent individuals and edges indicate spatial or social connections \cite{SkyrmsDynamicNetwork,ErezGraphs,SantosScaleFree,Ohtsuki,Taylor,SzaboFath,santos2008social,allen2014games}.  Edges can be weighted to indicate the strength of the connection. To study cooperation or other forms of social behavior, interactions can be modeled as matrix games.  Individuals play games with their neighbors, and the payoffs from these games determine reproductive success.  

Mathematical studies of evolutionary games on graphs \cite{Ohtsuki,Taylor,SzaboFath,taylor2011groups,chen2013sharp,cox2013voter,allen2014games,debarre2014social,durrett2014spatial,li2016evolutionary,pena2016evolutionary,Debarre2017fidelity} have typically assumed that the graph is regular, meaning that each individual has the same number of neighbors.  Recently, a condition was derived that determines which strategy is favored in any two-player, two-strategy game, on any weighted graph, under weak selection \cite{allen2017evolutionary,fotouhi2018conjoining}.  Weak selection means that the game has only a small effect on reproductive success. For nonweak selection, determining the outcome of evolutionary games on graphs is NP-hard \cite{ibsen2015computational}.

\begin{figure}[t!]
\begin{center}
\includegraphics{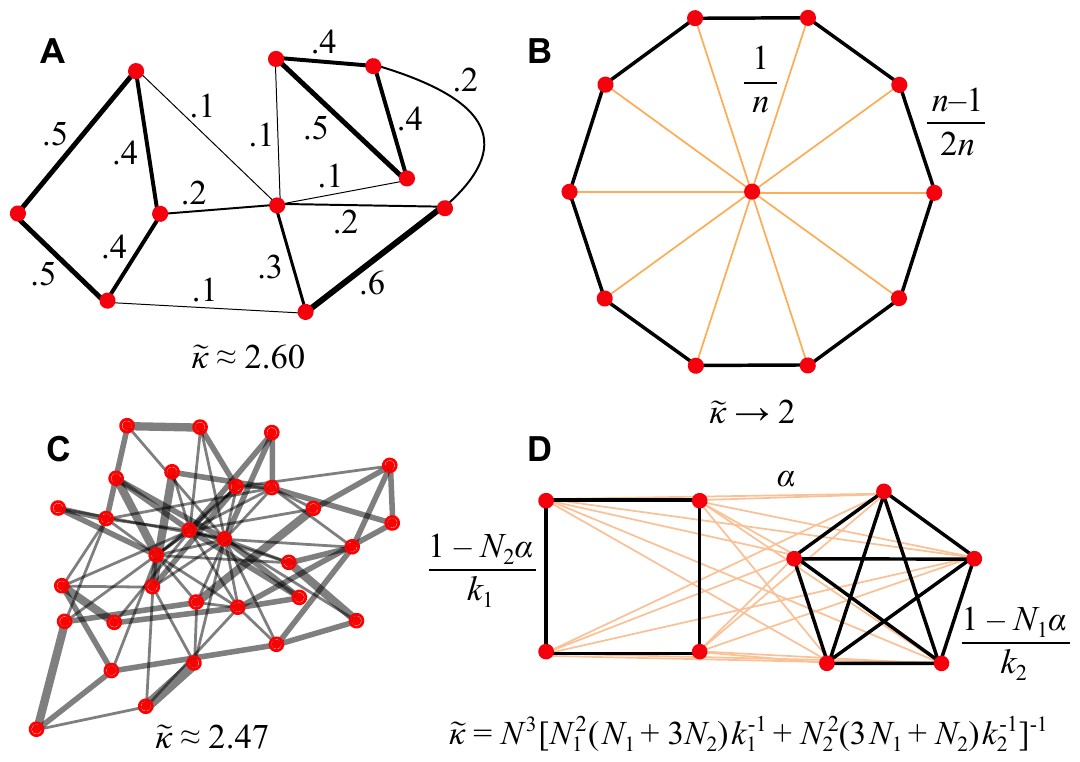}
\caption{\textbf{Isothermal graphs and their effective degrees.} 
A graph is isothermal if the sum of edge weights is the same for each vertex. 
The effective degree $\tilde{\kappa}$ of the graph, defined in Eq.~\eqref{eq:effdeg}, determines the outcome of evolutionary game dynamics. \textbf{(A)} An asymmetric isothermal graph; weights are shown for each edge. \textbf{(B)} A wheel graph, with one hub and $n$ wheel vertices. All connections with the hub have weight $1/n$. All connections in the periphery have weight $(n-1)/2n$.  As $n \to \infty$, the effective degree approaches 2. A formula for arbitrary $n$ is derived in Appendix \ref{app:wheel}.  \textbf{(C)} A 30-vertex graph generated with  preferential attachment \cite{barabasi1999emergence} and linking number $m=3$. Isothermal edge weights are obtained by quadratic programming (see Appendix \ref{app:numerical}).   The effective degree, $\tilde{\kappa} \approx 2.47$, is less than the average topological degree, $\bar{k} = 5.6$.  \textbf{(D)} An island model, with edges of weight $\alpha \ll 1$ between each inter-island pair of vertices. There are two islands: (i) a $k_1$-regular graph of size $N_1$ and (ii) a $k_2$-regular graph of size $N_2$. }
\label{fig:isothermal}
\end{center}
\end{figure}

A weighted graph is called \emph{isothermal} if the sum of edge weights is the same at each vertex (Fig.~\ref{fig:isothermal}). This property has a natural interpretation: suppose that the edge weights represent the amount of time that two individuals interact with each other.  Then the graph is isothermal as long as each individual devotes the same total time to interaction.  Importantly, some individuals may divide their time thinly among many contacts, while others focus their time on one or two contacts.

Isothermal graphs have special relevance for evolutionary dynamics. All vertices of an isothermal graph have the same reproductive value---meaning that each vertex contributes equally to the future population under neutral drift \cite{maciejewski2014reproductive,allen2015molecular}. The Isothermal Theorem \cite{ErezGraphs,pattni2015evolutionary} states that isothermal graphs neither amplify nor suppress the effects of  selection for mutations of constant fitness effect.

Here we analyze evolutionary games on isothermal graphs.  For this special class of graphs, we are able to obtain more powerful results than are available for arbitrary weighted graphs \cite{allen2017evolutionary,fotouhi2018conjoining}. We find that, for weak selection, the condition for success takes a particularly simple form, in which all effects of graph structure are captured in a single quantity---the \emph{effective degree}. An isothermal graph of effective degree $\tilde{\kappa}$ behaves like an unweighted $\tilde{\kappa}$-regular graph in its effect on strategy selection.  In particular, cooperation is favored on a large graph if and only if it provides a $\tilde{\kappa}$-fold benefit relative to the cost.  We derive bounds on $\tilde{\kappa}$ in terms of the graph's spectral gap (the difference between the two largest eigenvalues), establishing a link to the theory of expander graphs \cite{aldous2002reversible,hoory2006expander,lubotzky2012expander}.  Applying our results to power-law networks and to heterogeneous subdivided populations, we exhibit graphs of arbitrarily large average degree that provide arbitrarily strong support to cooperation.

\section*{Model}

We represent spatial structure by a weighted, connected, isothermal graph $G$ of size $N$.  The edge weight between vertices $i,j \in G$ is denoted $w_{ij}$. Without loss of generality, we scale edge weights so that $\sum_{j \in G} w_{ij} = 1$ for each vertex $i$.   In this way, edge weights may be interpreted as probabilities or frequencies of interaction. Edges are undirected, meaning $w_{ij}=w_{ji}$, and there are no self-loops: $w_{ii}=0$ for each $i$.   Two vertices are \emph{neighbors} if they are joined by an edge of positive weight; the number of neighbors of vertex $i$ is called its  \emph{topological degree} $k_i$.

\begin{figure}
\begin{center}
\includegraphics{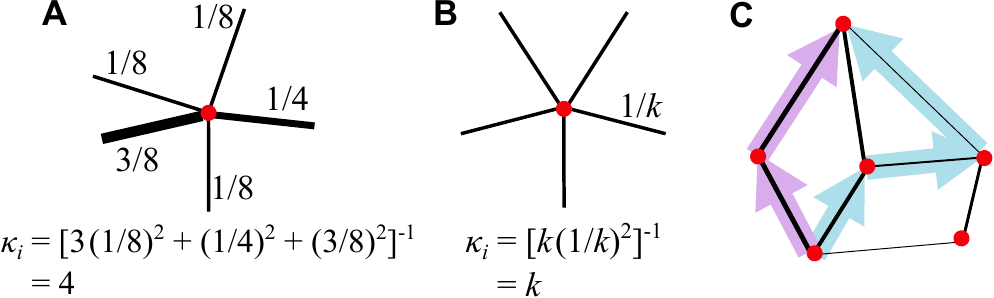}
\caption{\textbf{Simpson degree and remeeting time.} The Simpson degree $\kappa_i = \left( \sum_j w_{ij}^2 \right)^{-1}$ quantifies the effective number (or diversity) of neighbors of a vertex $i$, taking their edge weights into account. \textbf{A} If the edge weights to neighbors are nonuniform, the Simpson degree $\kappa_i$ is less than the topological degree $k_i$. Here, $\kappa_i=4$, which is less than the topological degree, $k_i=5$. \textbf{B} If each neighbor has equal edge weight $1/k$, the Simpson degree is equal to the topological degree, $k$. \textbf{C} The remeeting time $\tau_i$ is the expected time for two independent random walks from $i$ to meet each other. The effective degree $\tilde{\kappa}$ of a graph is the weighted harmonic average of the Simpson degrees, with weights given by the remeeting times.}
\label{fig:Simpson}
\end{center}
\end{figure}

Vertices in an isothermal graph may differ widely in the distribution of edge weights among their neighbors (Fig.~\ref{fig:isothermal}).  We quantify these differences using the \emph{Simpson degree} (Fig.~\ref{fig:Simpson}), a measure inspired by the Simpson index of biodiversity \cite{Simpson}. The Simpson degree $\kappa_i$ of vertex $i$ is defined as \cite{allen2013spatial,allen2014games}
\begin{equation}
\label{eq:Simpson}
\kappa_i = \left( \sum_{j \in G} w_{ij}^2 \right)^{-1}.
\end{equation}
In words, if individual $i$ randomly selects two neigbors, with probability proportional to edge weight, then $\kappa_i$ is the inverse probability that the same neighbor is selected twice.  The Simpson degree $\kappa_i$ quantifies the expected number of contacts of individual $i$, accounting for the time spent with each contact. If all edges from vertex $i$ have equal weight, then the Simpson and topological degrees are equal: $\kappa_i = k_i$.  Otherwise, $\kappa_i < k_i$, and $\kappa_i$ decreases as the distribution of edge weights from $i$ becomes more uneven.  

Individuals can be one of two types, A or B, corresponding to strategies in the game
\begin{equation}
\label{eq:generalgame}
\bordermatrix{
& \A & \B \cr
\A & a & b \cr
\B & c & d
}.
\end{equation}
Each time-step, each individual plays the game with all neighbors.  Payoffs from the game are translated into fecundity (reproductive rate) by $F_i = 1 + \delta f_i$, where $f_i$ is the edge-weighted average payoff that $i$ receives from neighbors, and $\delta$ is a parameter quantifying the strength of selection.  We study weak selection ($0 < \delta \ll 1$) as a perturbation of neutral drift ($\delta=0$). 

Evolution proceeds according to \emph{death-birth} updating \cite{Ohtsuki}.  First, a vertex $i \in G$ is chosen, with uniform probability, to be replaced.  A neighbor $j$ of $i$ is then chosen to reproduce, with probability proportional to $w_{ij} F_j$.  The offspring of $j$ replaces the occupant of $i$ and inherits the type of its parent.  Birth-death updating \cite{Ohtsuki} will be considered later.

Over time, one of the competing types will die out and the other will become fixed. Consider an initial state with a single vertex of type A chosen uniformly at random, and all other vertices of type B.  We define the fixation probability $\rho_\A$ as the (expected) probability that type A becomes fixed from this initial state. Similarly, $\rho_\B$ is the probability that type B becomes fixed from an initial state with one random (uniformly chosen) vertex of type B and all other vertices of type A.  We say A is favored if $\rho_\A > \rho_\B$.  

\section*{Results}

\subsection*{Condition for success}

We find that the key quantity characterizing an isothermal graph is its \emph{effective degree} $\tilde{\kappa}$, a weighted harmonic average of the graph's Simpson degrees: 
\begin{equation}
\label{eq:effdeg}
\tilde{\kappa} = \frac{\sum_i \tau_i}{\sum_i \tau_i \kappa_i^{-1}}.
\end{equation}
The weighting $\tau_i$ of vertex $i$ is the the expected remeeting time of two random walks that are initialized at $i$ (see Fig.~\ref{fig:Simpson}C and Appendix \ref{sec:coalescence}).  Remeeting times arise from tracing ancestries backwards in time as coalescing random walks \cite{kingman1982coalescent,aldous2002reversible,liggett2006interacting,WakeleyCoalescent}. The remeeting time $\tau_i$ is proportional to the probability that two neighbors competing to reproduce into vertex $i$ have different types. We observe that if all vertices have $k$ neighbors of equal weight, the effective degree is equal to the topological degree: $\tilde{\kappa}=k$.

We prove in Appendix \ref{app:DB} that strategy A is favored, for weak selection on an isothermal graph, if and only if
\begin{equation}
\label{eq:exactabcd}
\sigma a + b > c + \sigma d, \quad \text{with} \quad 
\sigma = \frac{\tilde{\kappa}+1-4\tilde{\kappa}/N}{\tilde{\kappa}-1}.
\end{equation}
In this condition, all effects of graph structure are captured in the effective degree $\tilde{\kappa}$.  Therefore, determining the conditions for success on a given isothermal graph amounts to computing the effective degree. This can be done in polynomial time by solving a system of linear equations for coalescence times (Appendix \ref{sec:coalescence}).  

As an interpretation of Condition (\ref{eq:exactabcd}), consider strategy A to represent cooperation and B to represent defection (noncooperation). Suppose that one's partner is equally likely to be either type. Then the expected reduction in one's payoff from playing A rather than B is $C=-\frac{1}{2}(a + b - c - d)$.  This quantity may be interpreted as the cost of cooperation.  The expected increase in one's \emph{partner's} payoff from playing A rather than B is $B=\frac{1}{2}(a - b + c - d)$, which may be interpreted as the benefit to the partner. Condition (\ref{eq:exactabcd}) then becomes 
\begin{equation}
\label{eq:critbc}
(N/\tilde{\kappa}-2)B > (N-2)C. 
\end{equation}
Suppose that $B,C>0$, meaning that cooperation is costly to the actor and beneficial to the recipient.  Then for $\tilde{\kappa} \ll N$, cooperation is favored as long as  $B/C > \tilde{\kappa}$. Well-known results for unweighted $k$-regular graphs \cite{Ohtsuki,Taylor,chen2013sharp,allen2014games} are recovered by substituting $k$ for $\tilde{\kappa}$. In contrast, if $\tilde{\kappa} \geq N/2$, then cooperation is never favored, but spiteful behaviors ($B<0, C>0$) can be favored.

\subsection*{Spectral gap bounds for expander graphs}

How does the effective degree $\tilde{\kappa}$ depend on other properties of the graph?  As a first observation, since $\tilde{\kappa}$ is a weighted average, it lies between the minimum and maximum Simpson degrees: $\kappa_{\min} \leq \tilde{\kappa} \leq \kappa_{\max}$.  Thus $(N/\kappa_{\min}-2)B > (N-2)C$ is necessary for cooperation to be favored, and $(N/\kappa_{\max}-2)B > (N-2)C$ is sufficient.

Stronger bounds on $\tilde{\kappa}$ can be obtained using the \emph{spectral gap} of the graph---the difference between the two largest eigenvalues of the adjacency matrix.  For isothermal graphs, the spectral gap is $g=1-\lambda_2$, where $\lambda_2$ is the second-largest eigenvalue. The spectral gap quantifies the expansion properties of a graph: how rapidly a ball grows in volume with respect to its radius, or how quickly a random walk ``forgets'' its initial position \cite{aldous2002reversible,hoory2006expander,lubotzky2012expander}. We define an  \emph{isothermal expander graph} as a large ($N \gg 1$) isothermal graph with non-negligible spectral gap ($g \centernot \ll 1$); see Appendix \ref{app:expander} for a formal definition using limits. Expander graphs have important applications in  mathematics \cite{lubotzky2012expander} and computer science \cite{hoory2006expander}. The spectral gap of an isothermal expander graph lies in the range $0<g\leq 1$.

We prove in Appendix \ref{app:tibounds} that remeeting times on an isothermal expander graph are asymptotically bounded by $\tau_i \leq N/g$, for each vertex $i$. We apply this result to bound the effective degree $\tilde{\kappa}$.  Let $\kA$ and $\kH$ denote the (unweighted) arithmetic and harmonic means, respectively, of the graph's Simpson degrees. Then the effective degree is bounded by 
\begin{equation}
\label{eq:specavgboundsMain}
g \kH \leq \tilde{\kappa} \leq \frac{\kA}{g}.
\end{equation}

Combining with \eqref{eq:critbc} we find that $B/C > g\kH$ is necessary for cooperation to be favored on an isothermal expander graph, and $B/C > \kA/g$ is sufficient.  Since the Simpson degree of a vertex cannot exceed its  topological degree, $B/C > \bar{k}/g$ also suffices for cooperation to be favored, where $\bar{k}=\frac{1}{N} \sum_i k_i$ is the arithmetic average topological degree.

Bounds (\ref{eq:specavgboundsMain}) can be improved by considering harmonic means over quantile ranges of the Simpson degree distribution.  For $0 \leq a < b \leq 1$, let $\kappa_{[a,b]}$ denote the harmonic mean of the Simpson degrees lying between the $a$th and $b$th quantiles.  For example, $\kappa_{[0,1/4]}$ denotes the harmonic mean over the smallest fourth (first quartile) of Simpson degrees.  The bounds in  \eqref{eq:specavgboundsMain} can then be sharpened to
\begin{equation}
\label{eq:quantilebounds}
\kappa_{[0,g]} \leq \tilde{\kappa} \leq \kappa_{[1-g,1]}.
\end{equation}
As $g \to 1$, the upper and lower bounds both converge to the harmonic mean Simpson degree $\kH$; thus $\tilde{\kappa}$ converges to $\kH$ as well in this limit.   For the case $g>1/2$, even tighter bounds are derived in Appendix \ref{app:sharper}.

\subsection*{Power-law networks}

\begin{figure}
\begin{center}
\begin{tabular}{cc}
\includegraphics[scale=0.7]{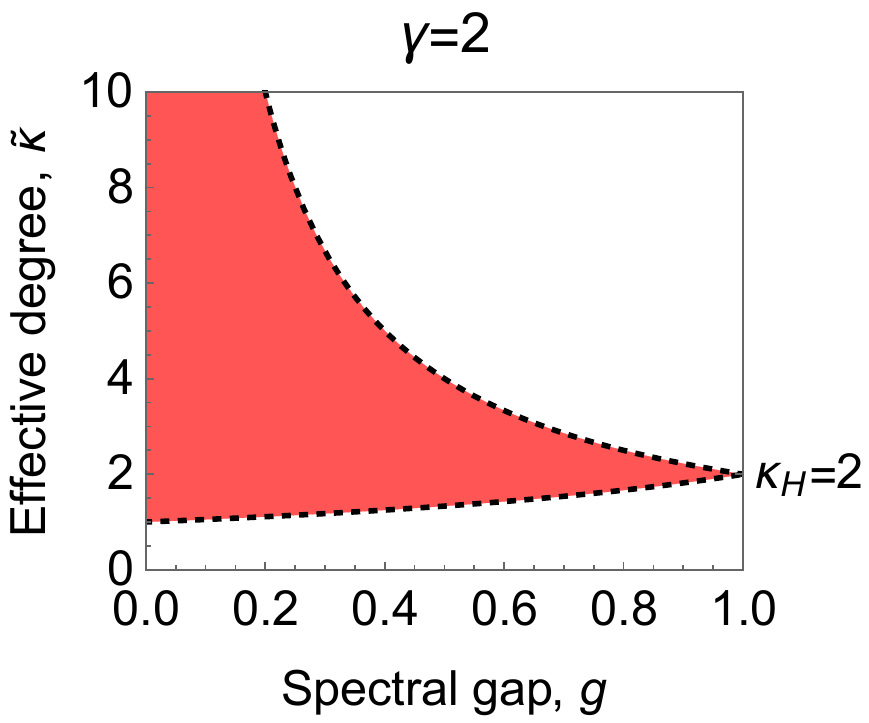} & \includegraphics[scale=0.7]{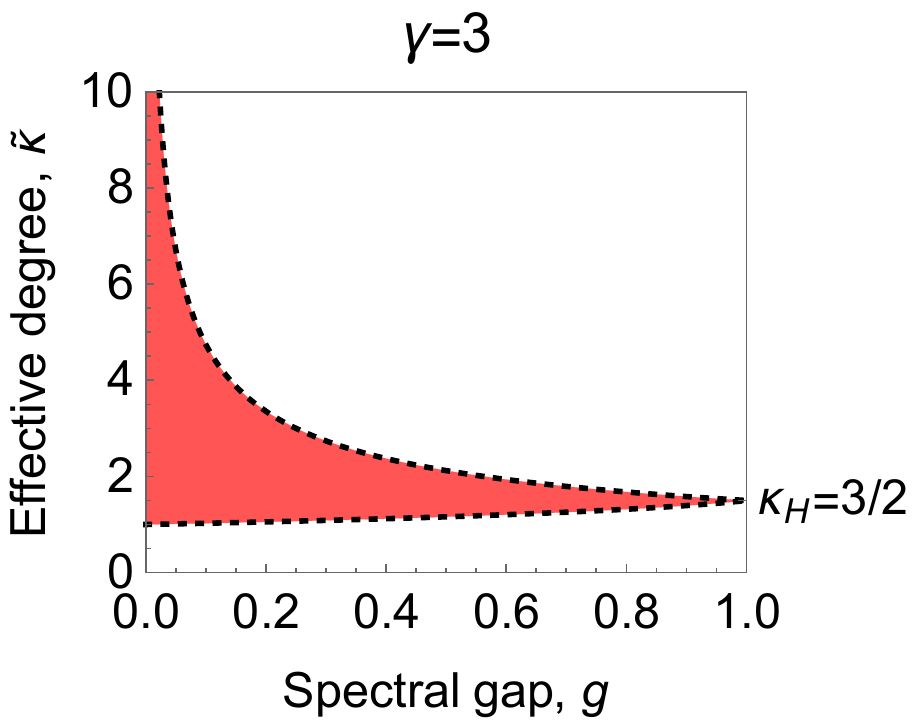}\\
\textsf{\textbf{A}} & \textsf{\textbf{B}}
\end{tabular}
\caption{\textbf{Bounds on effective degree for power-law expander graphs}. We consider a large isothermal graph for which the Simpson degree distribution is described by the density $f(\kappa) \propto \kappa^{-\gamma}$ on the range $[\kappa_0, \infty)$. The upper and lower bounds \eqref{eq:powerlawbounds1} are shown for \textbf{A} $\gamma=2$ and \textbf{B} $\gamma=3$, both with $\kappa_0=1$. As $g \to 1$, the upper and lower bounds both converge to the (unweighted) harmonic mean Simpson degree $\kappa_H$.}
\label{fig:powerlawbounds}
\end{center}
\end{figure}

\begin{figure}
\begin{center}
\begin{tabular}{cc}
\includegraphics[scale=0.8]{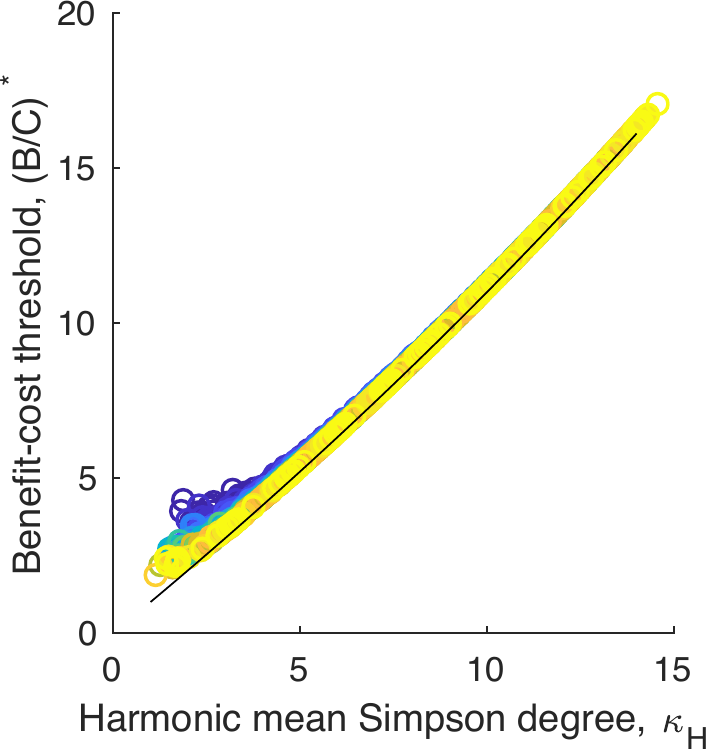} & \includegraphics[scale=0.8]{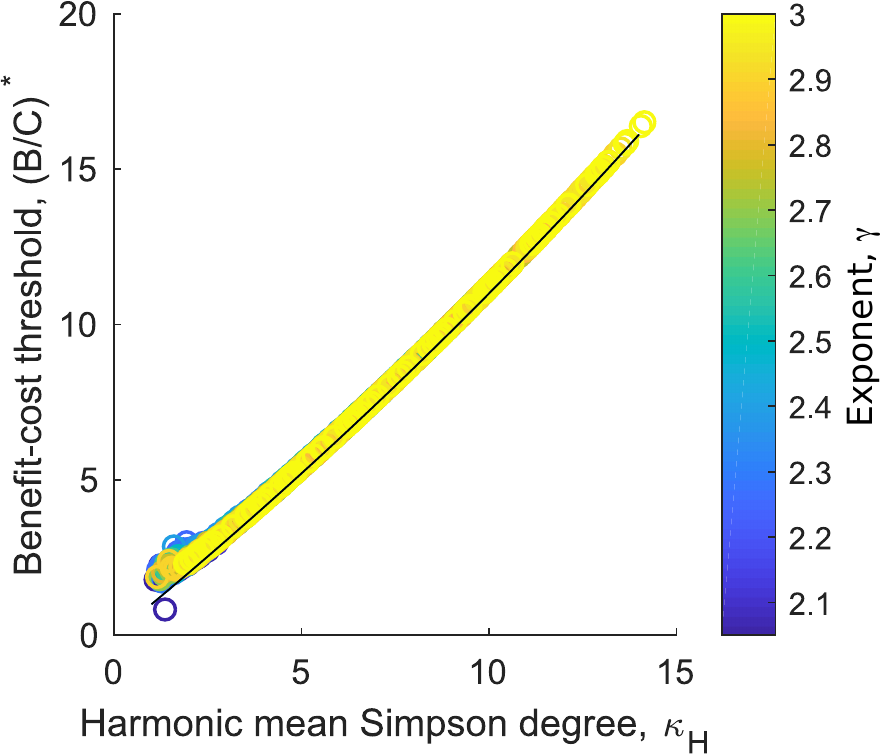}\\
\textsf{\textbf{A}} & \textsf{\textbf{B}}
\end{tabular}

\caption{\textbf{Conditions for cooperation on power-law networks.} The benefit-cost threshold $(B/C)^*= (N-2)/(N/ \tilde{\kappa} - 2)$ is plotted against the harmonic average Simpson degree $\kH$, for power-law networks generated by (A) shifted-linear preferential attachment \cite{dorogovtsev2000structure,krapivsky2001organization}, and (B) the configuration model.  Isothermal edge weights are generated via quadratic programming (see Appendix \ref{app:numerical}).  Each marker corresponds to a single graph. The approximation $(B/C)^* \approx (N-2)/(N/\kappa_H - 2)$ (solid line) works well for both models, with deviations only for $\gamma$ close to 2. }
\label{fig:random}
\end{center}
\end{figure}
 
 We apply our results to isothermal expander graphs with power-law degree distribution.  Let us hypothetically consider a  isothermal expander graph whose Simpson degree distribution is described by the power-law density $f(\kappa) \propto \kappa^{-\gamma}$, on the range $\kappa_0 \leq \kappa < \infty$, for arbitrary $\gamma \geq 2$ and $\kappa_0 \geq 1$. Evaluating \eqref{eq:quantilebounds} for the corresponding quantile function, $\kappa(x) = \kappa_0(1-x)^{-1/(\gamma-1)}$, yields
\begin{equation}
\label{eq:powerlawbounds1}
\left( \frac{\gamma}{\gamma-1} \right) \frac{\kappa_0 g}{1- (1-g)^{\gamma/(\gamma-1)}}  \leq \tilde{\kappa} \leq 
\left( \frac{\gamma}{\gamma-1} \right) \kappa_0 g^{-1/(\gamma-1)}.
\end{equation}
These bounds are illustrated in Figure \ref{fig:powerlawbounds}. Interestingly, for $\gamma=2$, the effective degree is bounded above by $2\kappa_0/g$, but the arithmetic average Simpson degree $\kA$ diverges to infinity.

To complement these analytical results, we numerically computed the effective degree $\tilde{\kappa}$ for random power-law networks generated by  preferential attachment and by the configuration model (Figs.~\ref{fig:isothermal}C, \ref{fig:random}).  We find that $\tilde{\kappa}$ is in most cases well-approximated by the harmonic mean Simpson degree $\kH$, which is simpler to compute.  It follows that the benefit-cost threshold for cooperation, $(B/C)^*= (N-2)/(N/ \tilde{\kappa} - 2)$, is well-approximated by $(N-2)/(N/\kH - 2)$. For large graphs, the condition for cooperation is approximately $B/C>\kH$. Note that this condition is easier to fulfill than either $B/C > \kA$ or $B/C > \bar{k}$, since $\kH \leq \kA \leq \bar{k}$, with both inequalities strict for non-regular graphs.

\subsection*{Island-structured populations}

Let us now consider a population divided into subpopulations (``islands'') with weak connections between them (Fig.~\ref{fig:isothermal}D).  The islands are represented by isothermal, vertex-transitive graphs, $G_1, \ldots, G_n$, which may differ in their size and structure.  An overall isothermal graph $G$ is formed by joining each inter-island pair of vertices by an edge of weight $\alpha \ll 1$, and rescaling intra-island edge weights correspondingly (see Appendix \ref{app:island}).  We prove that, if the island sizes are equal, the effective degree $\tilde{\kappa}$ of $G$ is the (unweighted) harmonic mean of the Simpson degrees $\kappa_1, \ldots, \kappa_n$ of the separate islands.  If the islands have different sizes, $\tilde{\kappa}$ is a weighted harmonic mean of $\kappa_1, \ldots, \kappa_n$, with weights depending only on the islands' sizes.

\begin{figure}
\begin{center}
\includegraphics{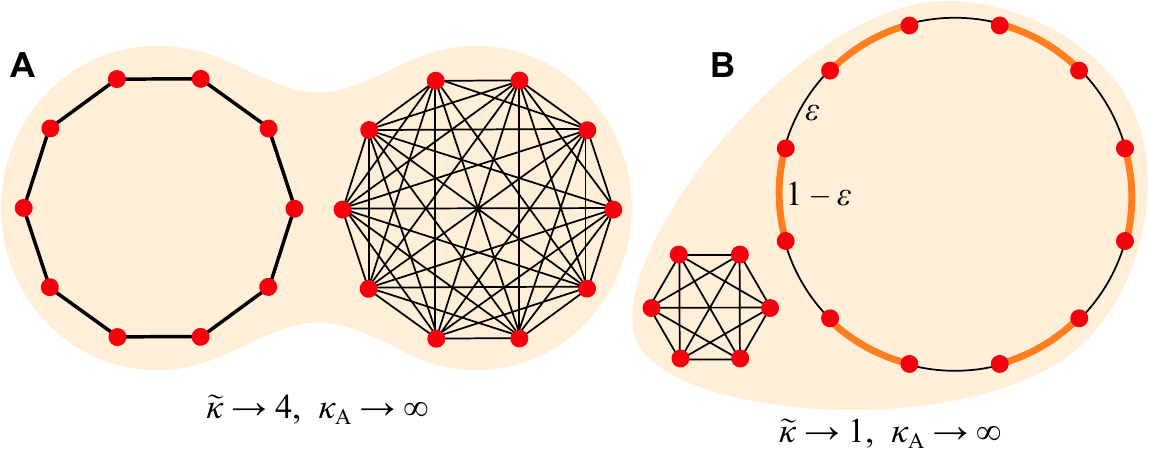}
\caption{\textbf{Island-structured ``super-promoters" of cooperation.} We use the island model (Fig.~\ref{fig:isothermal}D) to construct families of graphs whose effective degree $\tilde{\kappa}$ remains finite while the arithmetic average degree (either Simpson or topological) diverges to infinity.  \textbf{A} Two islands of equal size: a cycle and a complete graph.  As $N \to \infty$, the effective degree $\tilde{\kappa}$ converges to 4, which is the harmonic mean of 2 and infinity.  \textbf{B} A small complete graph and a large cycle with alternating edge weights, $\varepsilon \ll 1$ and $1-\varepsilon$.  Under the appropriate combination of limits, $\tilde{\kappa}$ converges to 1---meaning that all cooperative behaviors with $B>C>0$  are favored---while $\kA$ diverges.  Calculations are provided in Appendix \ref{app:island}.}
\label{fig:super}
\end{center}
\end{figure}

This result is significant for the evolution of cooperation. The harmonic mean of a set of numbers is dominated by its smallest elements.  Therefore, the overall effective degree $\tilde{\kappa}$ is most strongly influenced by the islands where the Simpson degree is smallest and cooperation is most favored.  This property allows us to design ``super-promoters" of cooperation (Fig. \ref{fig:super}), for which the effective degree $\tilde{\kappa}$ remains small---even as small as 1---while the arithmetic mean Simpson degree $\kA$ and topological degree $\bar{k}$ both diverge to infinity.

\subsection*{Birth-death updating}

We now turn to the birth-death update rule.  First, an individual $i$ is chosen, proportionally to its fecundity $F_i$, to reproduce.  The offspring of $i$ replaces neighbor $j$ with probability $w_{ij}$.  In this case, we show in Appendix \ref{app:BD} that type A is favored for weak selection if and only if 
\begin{equation}
\label{eq:exactBD}
\sigma a + b > c + \sigma d, \quad \text{with} \quad \sigma = (N-2)/N.
\end{equation}
This same condition for success has previously been derived for well-mixed populations, under a variety of update rules, with arbitrary selection strength and mutation rate \cite{kandori1993learning,NowakFinite,antal2009strategy}.  Here we have derived the same condition for birth-death updating on any isothermal graph, under weak selection. This result is reminiscent of the Isothermal Theorem \cite{ErezGraphs,pattni2015evolutionary}, which states that the fixation probability of a mutation of constant fitness, for birth-death updating on any isothermal graph, is the same as in a well-mixed population.  Condition (\ref{eq:exactBD}) extends this result to evolutionary games with weak selection.  

Rewriting Condition (\ref{eq:exactBD}) as $-(N-1)C-B>0$, we find that cooperation is never favored for positive $B$ and $C$. This generalizes, to all isothermal graphs, the previous finding that birth-death updating does not support cooperation on regular graphs \cite{Ohtsuki,Taylor,taylor2011groups,chen2013sharp,allen2014games,durrett2014spatial}. This effect arises because, under birth-death updating, any aid given to a neighbor increases the chances of being replaced by that neighbor.  Thus the benefits of cooperator assortment are exactly cancelled by local competition among cooperators \cite{TaylorViscous,taylor2011groups}. 

\section*{Discussion}

Here we have derived and analyzed conditions for evolutionary game success on weighted isothermal graphs, under weak selection.  Although there exists a general method for arbitrary weighted graphs  \cite{allen2017evolutionary,fotouhi2018conjoining}, our results for isothermal graphs are noteworthy in a number of respects.

First, the conditions for success take a simple form, \eqref{eq:exactabcd}, in which all effects of graph structure are captured in the effective degree $\tilde{\kappa}$.  For large graphs, cooperation is favored if it provides a $\tilde{\kappa}$-fold benefit relative to the cost. Since $\tilde{\kappa}$ is a weighted harmonic (rather than arithmetic) average, the presence of highly-connected hubs does not preclude support for cooperation. Indeed, $\tilde{\kappa}$ can be made arbitrarily close to 1---so that any cooperation producing a net benefit is favored---even as the average degree (topological or Simpson) diverges to infinity.

Second, we have derived explicit bounds for the effective degree---and, by extension, the benefit-cost threshold for cooperation---in terms of the graph's degree distribution and spectral gap.   Whether similar bounds can be obtained for arbitrary graphs is unknown.  The appearance of the spectral gap suggests an intriguing link between evolutionary game theory and the theory of expander graphs.  Currently, expansion properties are much better understood for regular graphs than for non-regular graphs \cite{aldous2002reversible,hoory2006expander,lubotzky2012expander}.  Isothermal graphs may serve as a useful class of intermediate generality for extending expansion theorems beyond regular graphs.

Third, we have shown that for birth-death updating on an isothermal graph, the conditions for $\rho_A > \rho_B$ under weak selection are entirely independent of the graph structure.  In particular, birth-death updating on isothermal graphs provides no support to the evolution of cooperation, because the benefits of cooperation are exactly cancelled by local competition. Such cancellation has been previously observed for regular graphs \cite{Ohtsuki,Taylor,taylor2011groups,chen2013sharp,allen2014games,durrett2014spatial} and other homogeneous population structures \cite{TaylorViscous,WilsonViscous}. Our result extends this finding to all isothermal graphs. Moreover, we find an unexpected link between such cancellation effects and the Isothermal Theorem \cite{ErezGraphs,pattni2015evolutionary}. The common thread is that, for birth-death updating on isothermal graphs, key aspects of the evolutionary process are invariant with respect to spatial structure.  Importantly, for non-isothermal graphs, Condition (\ref{eq:exactBD}) is not generally valid, and the conditions for success under birth-death updating vary from graph to graph \cite{Corina}. It therefore appears that the cancellation effects observed here and in previous work \cite{Ohtsuki,Taylor,taylor2011groups,chen2013sharp,allen2014games,durrett2014spatial} are restricted to isothermal graphs. The question of whether birth-death updating can ever support cooperation on a (non-isothermal) weighted graph remains open.

Our results add an important nuance to our understanding of the evolution of cooperation.  Previous work on regular graphs \cite{Ohtsuki,Taylor,SzaboFath,taylor2011groups,chen2013sharp,cox2013voter,allen2014games,debarre2014social,durrett2014spatial,li2016evolutionary,pena2016evolutionary,Debarre2017fidelity} showed that cooperation thrives when each individual has few neighbors, relative to the overall population size.  Condition (\ref{eq:exactabcd}) shows that it is not the raw number of neighbors that  matters, but their effective number, as quantified by $\tilde{\kappa}$.  Even in highly interconnected societies, cooperation can flourish if most individuals interact primarily with a few close contacts, rather than many loose acquaintances.

\section*{Acknowledgements}
B.~A.~is supported by National Science Foundation award \#DMS-1715315. We thank Babak Fotouhi for sharing data on birth-death updating.

\bibliographystyle{unsrt}

\appendix
\section{Model and notation}

We first review the model and introduce notation that will be used throughout this Appendix.  

\subsection{Isothermal graphs}

Population structure is described by a weighted isothermal graph $G$ with weights $w_{ij}$.  Without loss of generality we suppose $\sum_{j \in G} w_{ij}=1$ for each vertex $i$.  The graph is undirected ($w_{ij}=w_{ji}$ for each $i,j$) and has no self-loops ($w_{ii}=0$ for all $i$).  The Simpson degree of vertex $i$ is defined as 
\[
\kappa_i = \left(\sum_{j \in G} w_{ij}^2 \right)^{-1}.
\]

Random walks on $G$ are defined with step probabilities equal to edge weights: $p_{ij} = w_{ij}$ for each pair of vertices $i$ and $j$. The probability that an $n$-step random walk from $i$ terminates at $j$ is denoted $p_{ij}^{(n)}$. Note that
\[
p_{ii}^{(2)} = \sum_{j \in G} w_{ij}^2 = \kappa_i^{-1}.
\]
The stationary distribution for random walks on isothermal graphs is uniform: $\pi_i = 1/N$ for each $i \in G$.

The adjacency matrix of an isothermal graph $G$ is symmetric and doubly stochastic. It therefore has real eigenvalues $1=\lambda_1 \ge \lambda_2 \ge \ldots \ge \lambda_N$. Moreover, $\lambda_2 < 1$ as long as $G$ is connected.  The \emph{spectral gap} is defined as $g=1-\lambda_2$.  

\subsection{Evolutionary process}

The type occupying vertex $i$ is denoted $x_i\in \{0,1\}$, with 1 corresponding to A and 0 corresponding to B.  The population state is given by the vector $\vx = (x_i)_{i \in G} \in \{0,1\}^G$.

There are two competing types, A and B, corresponding to two strategies in the matrix game:
\begin{equation}
\label{eq:game}
\bordermatrix{ & \mathrm{A} & \mathrm{B}\cr \mathrm{A} & a & b\cr \mathrm{B} & c & d},
\end{equation}

In a given state, the edge-weighted average payoff to vertex $i$ is denoted $f_i = \sum_{j \in G} w_{ij} f_{ij}$, where $f_{ij}$ is the payoff that $i$ receives from interacting with $j$ according to the game.  

Payoff is translated into fecundity by  $F_i = 1+ \delta f_i$, where $\delta \geq 0$ quantifies the strength of selection.  The case $\delta = 0$ represents neutral drift, for which the game has no effect on selection.  Weak selection is the regime $0 < \delta \ll 1$.

Each time-step, an individual in a particular vertex $i$ is chosen to reproduce, and the offspring replaces the occupant of another vertex $j$. (We will use the shorthand ``$i$ replaces $j$" to describe such an event.)  The offspring inherits the type of the parent.  The probability that $i$ replaces $j$ in a given state depends on the specified update rule.  For death-birth updating, 
\[
\Prob[\text{$i$ replaces $j$}] = \frac{1}{N} \left( \frac{w_{ij} F_i}{\sum_k w_{kj} F_k} \right).
\]
For birth-death updating,
\[
\Prob[\text{$i$ replaces $j$}] = \left(\frac{F_i}{\sum_k F_k} \right) w_{ij}.
\]
Note that for neutral drift ($\delta=0$) the probability that $i$ replaces $j$ is $w_{ij}/N$ for both update rules.  This property is particular to isothermal graphs, and does not hold for the more general class of weighted undirected graphs.

There are two absorbing states: the state $\mathbf{1}$ for which only type A is present ($x_i=1\; \forall i \in G$), and the state $\mathbf{0}$ for which only type B is present ($x_i=0\; \forall i \in G$). All other states are transient \cite[Theorem 2]{allen2014measures}.  We define the \emph{fixation probability of A}, denoted $\rho_A$, as the expected probability of reaching state $\mathbf{1}$ from an initial state with a single A at a uniformly chosen random vertex, and all other vertices having type B. Likewise, we define the \emph{fixation probability of B}, denoted $\rho_B$, as the expected probability of reaching state $\mathbf{1}$ from an initial state with a single B at a uniformly chosen random vertex, and all other vertices having type A. 

\section{Derivation of conditions for success}

The conditions for a type to be favored on an arbitrary weighted graph, for weak selection, were derived in Refs.~\cite{allen2017evolutionary,allen2018mathematical}.  Here we provide a simplified derivation for the case of isothermal graphs.

\subsection{Weak selection analysis}

Define $\Delta(\vx)$ as the expected change in the number (absolute frequency) of A's from a given state $\vx$.  We compute this for the two update rules:
\begin{align*}
\Delta(\vx)  = 
\begin{cases}
\displaystyle \frac{1}{N} \sum_{j \in G} \left( -x_j +   \frac{\sum_{i \in G} x_i w_{ij} F_i}{\sum_{k \in G} w_{kj} F_k} \right) & \text{for Death-Birth}\\[6mm]
\displaystyle \frac{\sum_{i \in G} \left(x_i F_i - \sum_{j \in G} x_j w_{ij} \right)}{\sum_{k \in G} F_k} & \text{for Birth-Death}.
\end{cases}
\end{align*}

We observe that for neutral drift ($\delta=0$), $\Delta(\vx)=0$ for each state $\vx$.  For weak selection, we require the derivative of $\Delta'(\vx)=\frac{d \Delta(\vx)}{d\delta}|_{\delta=0}$. These derivatives can be expressed as
\begin{equation}
\Delta'(\vx) = \begin{cases}
\frac{1}{N^2} \sum_{i \in G} x_i \left(f_i - f_i^{(2)} \right) & \text{for Death-Birth}\\[2mm]
\frac{1}{N^2} \sum_{i \in G} x_i \left(f_i - f_i^{(1)} \right) & \text{for Birth-Death}.
\end{cases}
\end{equation}
Above, we have introduced the notation $f_i^{(n)} = \sum_j p_{ij}^{(n)} f_j$ for the expected payoff to the vertex at the end of an $n$-step random walk from $i$: 

We say that A is favored under weak selection if $\rho_A > \rho_B$ to first order in $\delta$.  Allen and McAvoy \cite{allen2018mathematical} showed this criterion can be evaluated by computing the expectation of $\Delta(\vx)$ over a particular probability distribution of population states $\vx$.  This distribution, called the \emph{neutral rare-mutation conditional distribution}, is obtained by (i) fixing $\delta=0$, (ii) introducing a mutation probability $u > 0$ to obtain a stationary distribution over states, (iii) conditioning this stationary distribution on both types being present, and (iv) taking the $u \to 0$ limit.  Denoting expectations over this distribution by $\langle \; \rangle$, Allen and McAvoy \cite{allen2018mathematical} showed that A is favored under weak selection if and only if 
\begin{equation}
\label{eq:Deltacondition}
\langle \Delta' \rangle > 0.
\end{equation}

\subsection{Coalescence times}
\label{sec:coalescence}

The conditions for success under weak selection can be expressed in terms of \emph{coalescence times}. Coalescence times are defined by considering a discrete-time process in which two random walkers start at vertices $i$ and $j$.  At each time, one of them is chosen, with equal probability, to take a step according to the usual step probabilities $p_{ij}$.  We denote the positions of the walkers at time $t$ by the pair $(X_t, Y_t)$, where $X_t$ and $Y_t$ are random variables with values in $G$.

Suppose that the two walkers start at vertices $i$ and $j$: $(X_0,Y_0) = (i,j)$. Let $M_{ij}$ denote the time until the walkers meet: $M_{ij}= \min \{t \ge 0: X(t)=Y(t)\}$.  The \emph{coalescence time} is defined as $\tau_{ij} = \E[M_{ij}]$.  These coalescence times satisfy the recurrence relation
\begin{equation}
\label{eq:taurecur}
\tau_{ij} = \begin{cases} 0 & i=j\\
1 + \frac{1}{2}  \sum_{k \in G} \left(w_{ik} \tau_{jk} + w_{jk} \tau_{ik} \right)& i \neq j.
\end{cases}
\end{equation}

For isothermal graphs, this coalescing random walk applies to both death-birth and birth-death updating.  (For arbitrary weighted graphs, a different coalescing random walk is needed for birth-death updating.) Duality between the the coalescing random walk and the neutral drift process \cite{holley1975ergodic} implies that all vertices $x_i, x_j, x_k$, and $x_\ell$, 
\begin{equation}
\label{eq:taux}
\langle x_i x_j - x_k x_\ell \rangle \; \propto \; \tau_{k \ell} - \tau_{ij}.
\end{equation}

Of particular interest is the \emph{remeeting time} for two walkers from the same vertex.  Suppose both walkers start at vertex $i$: $(X_0,Y_0) = (i,i)$.  Let $M_{ii}^+$ denote the first \emph{positive} time for which the walkers occupy the same vertex: $M_{ii}^+= \min \{t > 0: X(t)=Y(t)\}$.  We define the \emph{remeeting time} as $\tau_i = \E[M_{ii}^+]$. Remeeting times are related to coalescence times by
\begin{equation}
\label{eq:tauirecur}
\tau_i = 1 + \sum_{j\in G} w_{ij} \tau_{ij}.
\end{equation}
Remeeting times satisfy a return-time identity \cite{allen2017evolutionary}, which in the isothermal case is
\begin{equation}
\label{eq:Kac}
\sum_{i\in G} \tau_i = N^2.
\end{equation}

Another important quantity is the remeeting time $\tau^{(n)}$ from the two ends of an $n$-step random walk started from stationarity: 
\begin{equation}
\label{eq:taun}
\tau^{(n)} = \sum_{i,j \in G} p_{ij}^{(n)} \tau_{ij}.
\end{equation}  
The $\tau^{(n)}$ satisfy the recurrence relation
\begin{equation}
\label{eq:taunrecur}
\tau^{(n+1)} = \tau^{(n)} + \frac{1}{N}\sum_{i \in G}p_{ii}^{(n)} \tau_i - 1.
\end{equation}
Using Eqs.~\eqref{eq:Kac} and \eqref{eq:taunrecur}, and recalling the absence of self-loops ($p_{ii}^{(1)}=0$), we obtain
\begin{align}
\label{eq:t0}
\tau^{(0)} & = 0\\
\label{eq:t1}
\tau^{(1)} & =  N - 1\\
\label{eq:t2}
\tau^{(2)} & = N - 2\\
\label{eq:t3}
 \tau^{(3)} & = N + \frac{1}{N} \sum_{i \in G}\tau_i p_{ii}^{(2)} - 3.
\end{align}

Defining the effective degree $\tilde{\kappa}$ as
\[
\tilde{\kappa} = \frac{\sum_{i\in G} \tau_i}{\sum_{i\in G} \tau_i \kappa_i^{-1}} 
= N^2 \left ( \sum_i \tau_i \kappa_i^{-1} \right)^{-1} = N^2 \left ( \sum_i \tau_i p_{ii}^{(2)} \right)^{-1},
\]
we can rewrite Eq.~\eqref{eq:t3} as
\begin{equation}
\label{eq:t3kappa} \tau^{(3)} = N + N/\tilde{\kappa} - 3.
\end{equation}

\subsection{Conditions for success: Death-Birth}
\label{app:DB}

We will first consider the donation game
\begin{equation}
\label{eq:PD}
\bordermatrix{ & \mathrm{A} & \mathrm{B}\cr \mathrm{A} & b-c & -c\cr \mathrm{B} & b & 0},
\end{equation}
in which type A pays a cost $c$ to give a benefit $b$ to its partner. We will extend our results to the general game \eqref{eq:game} in Section \ref{sec:arbitrary}. For the donation game \eqref{eq:PD}, the payoff to vertex $i$ can be written
\begin{equation}
f_i = -cx_i + b x_i^{(1)},
\end{equation}
where $x_i^{(1)} = \sum_{j} w_{ij} x_j$ is the average type among neighbors of $i$.

To determine the condition for A to be favored for death-birth updating, we compute
\begin{align*}
\langle \Delta' \rangle & = \frac{1}{N^2} \sum_{i \in G} \left \langle x_i \left(f_i - f_i^{(2)} \right) \right \rangle\\
& = \frac{1}{N^2} \sum_{i \in G} \left ( -c \left \langle x_i \left(x_i-  x_i^{(2)} \right)\right \rangle 
+ b\left \langle x_i \left(x_i^{(1)}-  x_i^{(3)} \right)\right \rangle  \right).
\end{align*}
Applying Eqs.~\eqref{eq:Deltacondition}, \eqref{eq:taux} and \eqref{eq:taun}, we obtain the condition
\begin{equation}
-c \tau^{(2)} + b \left (\tau^{(3)} - \tau^{(1)} \right) > 0.
\end{equation}
Now using Eqs.~\eqref{eq:t1}--\eqref{eq:t3}, we find that A is favored under weak selection if and only if 
\begin{equation}
\label{eq:DBcond}
-c (N-2) + b \left (N/\tilde{\kappa}-2 \right) > 0.
\end{equation}

\subsection{Conditions for success: Birth-Death}
\label{app:BD}

For birth-death updating, we compute
\begin{align*}
\langle \Delta' \rangle & = \frac{1}{N^2} \sum_{i \in G} \left \langle x_i \left(f_i - f_i^{(1)} \right) \right \rangle\\
& = \frac{1}{N^2} \sum_{i \in G} \left ( -c \left \langle x_i \left(x_i-  x_i^{(1)} \right)\right \rangle 
+ b\left \langle x_i \left(x_i^{(1)}-  x_i^{(2)} \right)\right \rangle  \right).
\end{align*}
Applying Eqs.~\eqref{eq:Deltacondition}, \eqref{eq:taux} and \eqref{eq:taun} yields the condition
\begin{equation}
-c \tau^{(1)} + b \left (\tau^{(2)} - \tau^{(1)} \right) > 0.
\end{equation}
Substituting from Eqs.~\eqref{eq:t1}--\eqref{eq:t3}, we obtain that A is favored under weak selection if and only if 
\begin{equation}
\label{eq:BDcond}
-c (N-1) - b > 0.
\end{equation}

\subsection{Extension to arbitrary $2 \times 2$ games}
\label{sec:arbitrary}

We turn now to the general $2 \times 2$ game \eqref{eq:game}.  The Structure Coefficient Theorem \cite{Corina} states that, for a general class of evolutionary game theory processes, including the process considered here, the condition for success under weak selection takes the form
\begin{equation}
\label{eq:sigma}
\sigma a + b > c + \sigma d,
\end{equation}
for some structure coefficient $\sigma$ that is independent of the game matrix.
Let us set $C = -(a + b - c - d)$ and $B = a - b + c - d$.  Note that, for the donation game, $C=c$ and $B=b$.   Condition \eqref{eq:sigma} now becomes equivalent to
\begin{equation}
\label{eq:sigma2}
-(\sigma + 1) C + (\sigma - 1)B > 0.
\end{equation}
It follows that the condition for success must be expressible in the form 
\[
-CK_1 + BK_2>0,
\] 
where $K_1$ and $K_2$ are independent of the game matrix and are unique up to a common positive multiple. We may therefore replace $c$ and $b$ by $C$ and $B$, respectively, in Conditions \eqref{eq:DBcond} and \eqref{eq:BDcond}.

\section{Bounds on remeeting times}

Here we derive upper and lower bounds on the remeeting time $\tau_i$ from a single vertex, in terms of the spectral gap $g$.  These will be used to obtain bounds on the effective degree $\tilde{\kappa}$.

\subsection{Background on hitting times}

For a single random walk on $G$, let $h_{ij}$ denote the expected hitting time to vertex $j$ when starting from vertex $i$.  These hitting times satisfy the recurrence equations
\begin{equation}
\label{eq:hrecur}
\begin{cases}
h_{ij} = 1 + \sum_k w_{ik} h_{kj} & \text{for $i \neq j$,}\\
h_{ii} = 0 & \text{for all $i$}.
\end{cases}
\end{equation}
Since the stationary distribution on isothermal graphs is uniform, it follows from the return-time identity (e.g. \cite[Lemma 2.5]{aldous2002reversible}) that the expected time for a random walk to return to its initial vertex is $N$, regardless of which initial vertex is used.  Combining with Eq.~\eqref{eq:hrecur}, we have the identity
\begin{equation}
\label{eq:hreturn}
1 + \sum_k w_{ik} h_{ik} = N.
\end{equation}

Let $h_{* j} = \frac{1}{N} \sum_i h_{ij}$ denote the expected hitting time to $j$ from a vertex chosen uniformly at random.  Corollary 3.14 of Ref.~\cite{aldous2002reversible} gives the identity
\begin{equation}
\label{eq:hdiff}
h_{ij} - h_{* j}  =  h_{ji} - h_{* i}.
\end{equation}

Proposition 3.17 of Ref.~\cite{aldous2002reversible} gives bounds on $h_{* j}$ in terms of the spectral gap.  In the case of an isothermal graph with no self-loops, these bounds are
\begin{equation}
\label{eq:hittingbounds}
\frac{(N-1)^2}{N} \leq h_{* j} \leq \frac{N-1}{g}.
\end{equation}

\subsection{Spectral gap bounds on remeeting times}
\label{app:tibounds}

We apply the bound \eqref{eq:hittingbounds} to obtain bounds on the remeeting times $\tau_i$.  

\begin{theorem}
\label{lem:taubounds}
For each $i \in G$,
\begin{equation}
\label{eq:taubounds}
2N - \frac{N-1}{g} - \frac{2N-1}{N} \leq \tau_i \leq \frac{N-1}{g} + \frac{2N-1}{N}.
\end{equation}
\end{theorem}
We note that the lower bound in Eq.~\eqref{eq:taubounds} is not necessarily positive.

\begin{proof}
Our proof is a variation on the proof of Proposition 14.5 of Ref.~\cite{aldous2002reversible}. Consider the process of two random walkers $(X_t, Y_t)_{t=0}^\infty$ described in Section \ref{sec:coalescence}, where both walkers start at vertex $i$: $(X_0, Y_0)=(i,i)$.  We define the real-valued stochastic process $(S_t)_{t=0}^\infty$ by
\begin{equation}
\label{eq:Stdef}
S_t = \begin{cases}  N - h_{* i} & t=0\\
t + h_{X_tY_t} -h_{* Y_t} & t \geq 1
\end{cases}
\end{equation}
Eqs.~\eqref{eq:hrecur}, \eqref{eq:hreturn}, and \eqref{eq:hdiff} imply that $S_t$ for $0 \le t \le M_{ii}^+$ is a Martingale.  We then have
\begin{align*}
N - h_{* i} & = S_0\\
 & = \E[S_{M_{ii}^+}] & \text{by the Optional Stopping Theorem \cite{doob1953stochastic}}\\
& = \E[M_{ii}^+] - \E \left[ h_{* Y_{M_{ii}^+}} \right] & \text{by the construction of $S_t$}\\
& = \tau_i - \E \left[ h_{* Y_{M_{ii}^+}} \right] & \text{by definition of $\tau_i$.}
\end{align*}
Rearranging,
\begin{equation}
\label{eq:martingaleiso}
\tau_i =  N - h_{* i} +  \E \left[ h_{* Y_{M_{ii}^+}} \right].
\end{equation}
The upper bound on $\tau_i$ is obtained by applying the lower bound of \eqref{eq:hittingbounds} to the second term of Eq.~\eqref{eq:martingaleiso}, and the upper bound of \eqref{eq:hittingbounds} to the third term of Eq.~\eqref{eq:martingaleiso}.  Applying these bounds in the opposite fashion gives the lower bound on $\tau_i$.
\end{proof}

 We note that the average of the upper and lower bounds in \eqref{eq:taubounds} is $N$, which is equal to the average value of $\tau_i$ according to Eq.~\eqref{eq:Kac}.  We also observe that as $N \to \infty$, the lower and upper bounds are asymptotically $N(2-g^{-1}) + \mathcal{O}(1)$ and $Ng^{-1} + \mathcal{O}(1)$, respectively.  Thus, in the $N \to \infty$ limit, the lower bound is relevant (i.e.,~positive) for $g>1/2$.

\section{Bounds on the effective degree}

\subsection{Quantile bounds for an arbitrary isothermal graph}

Recall that $\tilde{\kappa}$ is a weighted harmonic average of the Simpson degrees $\kappa_i$. The weights $\tau_i$ are bounded by Eq.~\eqref{eq:taubounds}.  Upper and lower bounds on $\tilde{\kappa}$ can therefore be obtained by placing the maximum weight on vertices that have the largest or smallest Simpson degrees, respectively, taking into account that the sum of the $\tau_i$ is constrained by Eq.~\eqref{eq:Kac}.

To formalize this idea, we introduce the quantity 
\begin{equation}
\label{eq:ghat}
\hat{g}= \left( \frac{N-1}{Ng} + \frac{2N-1}{N^2} \right)^{-1},
\end{equation}
so that the bounds \eqref{eq:taubounds} become
\begin{equation}
\label{eq:taughat}
N(2-\hat{g}^{-1}) \leq \tau_i \leq N\hat{g}^{-1}.   
\end{equation}
We note that $\hat{g}=g+\mathcal{O}(N^{-1})$ as $N \to \infty$, and also that $\hat{g} > g$ as long as $g<1/2$.

We write the effective degree of $G$ in the form
\begin{equation}
\label{eq:kappa}
\tilde{\kappa} = \left( \sum_{i \in G} \left( \frac{\tau_i}{N^2} \right) \kappa_i^{-1} \right)^{-1}.
\end{equation}
By Eqs.~\eqref{eq:Kac} and \eqref{eq:taubounds}, the weights, $\tau_i/N^2$, are subject to the constraints
\begin{equation}
\label{eq:tauconstraints}
\frac{\tau_i}{N^2} \leq \frac{1}{N\hat{g}}, \qquad \sum_{i \in G} \frac{\tau_i}{N^2} = 1.
\end{equation}

Next we index the vertices in order of increasing Simpson degree, so that
\begin{equation}
\label{eq:kappasequence}
\kappa_1 \leq \kappa_2 \leq \dots \leq \kappa_N.
\end{equation}
We define $\Harmonic_{[0,\hat{g}]}[\kappa]$ to be the harmonic average over the fraction $\hat{g}$ of vertices with the smallest Simpson degree:
\begin{equation}
\label{eq:Hlower}
\Harmonic_{[0,\hat{g}]}[\kappa] = \left( \frac{1}{ N\hat{g} } \sum_{i=1}^{\lfloor N\hat{g} \rfloor} \kappa_i^{-1} + \left(1- \frac{\lfloor N\hat{g} \rfloor}{N\hat{g}}  \right) \kappa_{\lfloor N\hat{g} \rfloor + 1}^{-1} \right)^{-1}.
\end{equation}
Above, $\lfloor N\hat{g} \rfloor$ denotes the greatest integer less than or equal to $N\hat{g}$ (i.e. the floor function of $N\hat{g}$).  In Eq.~\eqref{eq:Hlower}, the $\lfloor N\hat{g} \rfloor$ vertices with the smallest Simpson degree are each given weight $1/(N\hat{g})$. The remainder of the weight is placed on the vertex with next-smallest Simpson degree $\kappa_{\lfloor N\hat{g} \rfloor + 1}$, so that the total weights sum to one.

Similarly, we define  $\Harmonic_{[1-\hat{g},1]}[\kappa]$ to be the harmonic average over the fraction $\hat{g}$ of vertices with the largest Simpson degree:
\begin{equation}
\label{eq:Hupper}
\Harmonic_{[1-\hat{g},1]}[\kappa] = \left( \left(1- \frac{\lfloor N\hat{g} \rfloor}{N\hat{g}}  \right) \kappa_{N-\lfloor N\hat{g} \rfloor}^{-1} + \frac{1}{ N\hat{g} } \sum_{i=N-\lfloor N\hat{g} \rfloor+1}^{N} \kappa_i^{-1}  \right)^{-1}.
\end{equation}

$\Harmonic_{[0,\hat{g}]}[\kappa]$ and $\Harmonic_{[1-\hat{g},1]}[\kappa]$ represent the smallest and largest values, respectively, of the right-hand side of Eq.~\eqref{eq:kappa} that are achievable under the constraints \eqref{eq:tauconstraints}.  We therefore have
\begin{equation}
\label{eq:finitespecavgbounds}
\Harmonic_{[0,\hat{g}]}[\kappa] \leq \tilde{\kappa} \leq \Harmonic_{[1-\hat{g},1]}[\kappa].
\end{equation}

\subsection{Isothermal expander graphs}
\label{app:expander}

We now consider a sequence of isothermal graphs $\{G_j\}_{j=1}^\infty$ with corresponding sizes $N_j$ and spectral gaps $g_j$.  We define this to be a sequence of isothermal \emph{expander} graphs if $\lim_{j \to \infty} N_j = \infty$ and $\liminf_{j \to \infty} g_j > 0$.  By passing to a subsequence if necessary, we can assume $\lim_{j \to \infty} g_j = g>0$. Notice that this also entails $\lim_{N \to \infty} \hat{g}_j = g$. 

To obtain limiting values of the bounds \eqref{eq:finitespecavgbounds} for such a sequence, we turn the degree sequence \eqref{eq:kappasequence} for graph $G_j$ into a quantile function---a nondecreasing piecewise-constant function $\kappa_j(x)$, defined for $0 < x \leq 1$.  Let $\kappa_{i,j}$ denote the $i$th smallest Simpson degree among the vertices of $G_j$.  We then define $\kappa_j(x) = \kappa_{ \lceil N_j x \rceil, j}$, where $\lceil \; \rceil$ denotes the ceiling function.  Eqs.~\eqref{eq:Hlower} and \eqref{eq:Hupper} can then be rewritten as 
\begin{align}
\label{eq:Hintlower}
\Harmonic_{[0,\hat{g}_j]}[\kappa_j] & =  \left( \hat{g}_j^{-1} \int_0^{\hat{g}_j} \big(\kappa_j(x)\big)^{-1}\; dx \right)^{-1}\\
\label{eq:Hintupper}
\Harmonic_{[1-\hat{g}_j,1]}[\kappa_j] & = \left( \hat{g}_j^{-1} \int_{1-\hat{g}_j}^1 \big(\kappa_j(x)\big)^{-1}\; dx \right)^{-1}.
\end{align}

Now suppose that as $j \to \infty$, $\kappa_j(x)$ converges pointwise to some real-valued function $\kappa(x)$ on the interval $[0,1)$.  Then $\kappa(x)$ is the quantile function of the limiting Simpson degree distribution. That is, a fraction $x$ of the Simpson degrees lie below $\kappa(x)$, for each $0 \leq x < 1$.  We allow for the possibility that $\lim_{x \to 1} \kappa(x)=\infty$.

As $j \to \infty$, the integrals in Eqs.~\eqref{eq:Hintlower}--\eqref{eq:Hintupper} converge to
\begin{equation}
\label{eq:specavgbounds}
\Harmonic_{[0,g]}[\kappa] \leq \tilde{\kappa} \leq \Harmonic_{[1-g,1]}[\kappa],
\end{equation}
with the limiting bounds given by
\begin{align}
\Harmonic_{[0,g]}[\kappa] & =  \left( g^{-1} \int_0^g \big(\kappa(x)\big)^{-1}\; dx \right)^{-1}\\
\Harmonic_{[1-g,1]}[\kappa] & = \left( g^{-1} \int_{1-g}^1 \big(\kappa(x)\big)^{-1}\; dx \right)^{-1}.
\end{align}
Convergence is guaranteed by the Dominated Convergence Theorem, using the fact that $0<\big(\kappa_j(x)\big)^{-1} \leq 1$ for all $j\geq 1$ and all $x \in [0,1)$.

\subsection{Arithmetic and harmonic mean bounds}

We can also obtain simpler, but looser, bounds that depend only on the arithmetic and harmonic mean Simpson degree. We begin with a single isothermal graph $G$. For a lower bound, using inequality \eqref{eq:taughat}, we have
\begin{equation}
\label{eq:glower}
\tilde{\kappa} = \left( \sum_{i \in G} \frac{\tau_i}{N^2} \kappa_i^{-1} \right)^{-1}
\geq \left( \frac{1}{N\hat{g}} \sum_{i \in G} \kappa_i^{-1} \right)^{-1} = \hat{g} \kH,
\end{equation}
where $\kH$ denotes the (unweighted) harmonic average Simpson degree.

For the upper bound, we have the following series of inequalities:
\begin{equation}
\label{eq:meansinequalities}
\Harmonic_{[1-\hat{g},1]}[\kappa] \leq \operatorname{A}_{[1-\hat{g},1]}[\kappa] \leq \frac{1}{\hat{g}} \kA.
\end{equation}
Above, $\operatorname{A}_{[1-\hat{g},1]}[\kappa]$ is the arithmetic average of the fraction $\hat{g}$ of Simpson degrees that are the largest, defined similarly to $\Harmonic_{[1-\hat{g},1]}[\kappa]$.  $\kA=\operatorname{A}_{[0,1]}[\kappa]$ is the average Simpson degree over all vertices. The first inequality in \eqref{eq:meansinequalities} is the arithmetic-harmonic means inequality, while the second reflects the fact that $\frac{1}{\hat{g}} \kA$ can be obtained by adding additional positive terms to the sum in $\operatorname{A}_{[1-\hat{g},1]}[\kappa]$.  

Combining \eqref{eq:specavgbounds}, \eqref{eq:glower}, and \eqref{eq:meansinequalities}, we have 
\begin{equation}
\label{eq:ghatbounds}
\hat{g}\kH \leq \tilde{\kappa} \leq \frac{\kA}{\hat{g}} .
\end{equation}

Turning now to a sequence of isothermal expander graphs $\{G_j\}_{j=1}^\infty$, the bounds \eqref{eq:ghatbounds} converge to
\begin{equation}
\label{eq:gbounds}
g\kH \leq \tilde{\kappa} \leq \frac{\kA}{g}.
\end{equation}
We note that, for any given graph with spectral gap $g<1/2$, the bounds \eqref{eq:ghatbounds} are stronger than \eqref{eq:gbounds}, since $\hat{g}>g$ in this case.  We also note that $\kA$ may diverge as $j \to \infty$; thus the upper bound in \eqref{eq:gbounds} may be infinite.

\subsection{Sharper bounds for large spectral gap}
\label{app:sharper}

In the case that the lower bound on $\tau_i$ in \eqref{eq:taughat} is positive, we can further sharpen the bounds on the effective degree by assigning one bound in \eqref{eq:taughat} to the lower half of Simpson degrees and the other bound to the upper half.  More precisely, suppose the vertices are ordered as in \eqref{eq:kappasequence}.  Then a lower bound is given by 
\begin{equation}
\tilde{\kappa} \geq 
\left( \frac{\hat{g}^{-1}}{N} \sum_{i=1}^{N/2} \kappa_i^{-1} + \frac{2-\hat{g}^{-1}}{N}  \sum_{i=N/2+1}^N \kappa_i^{-1} \right)^{-1},
\end{equation}
if $N$ is even, and 
\begin{equation}
\tilde{\kappa} \geq 
\left( \frac{\hat{g}^{-1}}{N} \sum_{i=1}^{(N-1)/2} \kappa_i^{-1} + \frac{\kappa_{(N+1)/2}^{-1} }{N} + \frac{2-\hat{g}^{-1}}{N}  \sum_{i=(N+3)/2}^N \kappa_i^{-1} \right)^{-1}.
\end{equation}
if $N$ is odd.  Similarly, for an upper bound, we have
\begin{equation}
\tilde{\kappa} \leq 
\left( \frac{2-\hat{g}^{-1}}{N} \sum_{i=1}^{N/2} \kappa_i^{-1} + \frac{\hat{g}^{-1}}{N}   \sum_{i=N/2+1}^N \kappa_i^{-1} \right)^{-1},
\end{equation}
if $N$ is even, and 
\begin{equation}
\tilde{\kappa} \leq 
\left( \frac{2-\hat{g}^{-1}}{N} \sum_{i=1}^{(N-1)/2} \kappa_i^{-1} + \frac{\kappa_{(N+1)/2}^{-1} }{N} + \frac{\hat{g}^{-1}}{N}   \sum_{i=(N+3)/2}^N \kappa_i^{-1} \right)^{-1}.
\end{equation}
if $N$ is odd.

If we now consider a family of isothermal expander graphs $\{G_j\}_{j=1}^\infty$ as in the previous section, with limiting spectral gap $g>0$ and limiting degree sequence described by the real-valued function $\kappa(x)$, the corresponding bounds on the effective degree converge to 
\begin{equation}
\label{eq:largegbounds}
\left(  \frac{1}{2g} \klow^{-1} 
+ \left(1-\frac{1}{2g} \right) \khigh^{-1}  \right)^{-1} \leq \tilde{\kappa} 
\leq \left( \left(1-\frac{1}{2g} \right) \klow^{-1} 
+ \frac{1}{2g} \khigh^{-1}  \right)^{-1},
\end{equation}
where $\klow=\Harmonic_{\left[0,\frac{1}{2}\right]}[\kappa]$ and $\khigh=\Harmonic_{\left[\frac{1}{2},1\right]}[\kappa]$ are the harmonic averages of the smaller and larger half of Simpson degrees, respectively.  Bounds \eqref{eq:largegbounds} can also be written as
\begin{equation}
\label{eq:largegbounds2}
\frac{2g \klow \khigh}
{(2g-1) \klow + \khigh}
\leq \tilde{\kappa}  \leq 
\frac{2g \klow \khigh}
{\klow + (2g-1) \khigh}.
\end{equation}

\section{Power-law graphs}

Let us now suppose that, for a family of isothermal expander graphs with limiting spectral gap $g>0$, the Simpson degree distribution converges to the power-law density $f(\kappa) = (\gamma - 1) \kappa_0^{\gamma-1} \kappa^{-\gamma}$, valid for $\kappa \in [\kappa_0,\infty)$, with exponent $\gamma \geq 2$. 

\subsection{Quantile bounds}

To apply the bounds \eqref{eq:specavgbounds}, we must determine the quantile function $\kappa(x)$.  For this we solve the equation
\begin{align*}
x & = (\gamma - 1)\kappa_0^{\gamma-1} \int_{\kappa_0}^y \kappa^{-\gamma} \; d\kappa\\
& = 1-(y/\kappa_0)^{1-\gamma}.
\end{align*} 
Solving for $y$ yields the quantile function:
\begin{equation}
\kappa(x) = y = \kappa_0 (1-x)^{1/(1-\gamma)}.
\end{equation}

Now we calculate
\begin{equation}
\label{eq:gammalower}
\begin{split}
\Harmonic_{[0,g]}[\kappa] & = \left( g^{-1} \int_{\kappa_0}^{\kappa(g)} f(\kappa) \kappa^{-1} d\kappa \right)^{-1}\\
& = \left(g^{-1} (\gamma-1) \kappa_0^{\gamma-1} \int_{\kappa_0}^{\kappa_0 (1-g)^{1/(1-\gamma)}} \kappa^{-\gamma-1} \; d\kappa \right)^{-1}\\
& = \left( \frac{\kappa_0 \gamma}{\gamma-1} \right) \frac{ g}{1- (1-g)^{\gamma/(\gamma-1)}}.
\end{split}
\end{equation}
Similarly,
\begin{equation}
\label{eq:gammaupper}
\begin{split}
\Harmonic_{[1-g,1]}[\kappa] 
& = \left( g^{-1} \int_{\kappa(1-g)}^\infty f(\kappa) \kappa^{-1} \right)^{-1}\\
& = \left( g^{-1}  (\gamma-1) \kappa_0^{\gamma-1} \int_{\kappa_0 g^{1/(1-\gamma)}}^\infty \kappa^{-\gamma-1} \right)^{-1}\\
& = \frac{\kappa_0 \gamma}{\gamma-1} g^{-1/(\gamma-1)}.
\end{split}
\end{equation}
Thus we have the bounds
\begin{equation}
\label{eq:powerlawbounds}
\left( \frac{ \gamma}{\gamma-1} \right) \frac{\kappa_0 g}{1- (1-g)^{\gamma/(\gamma-1)}}  \leq \left( \frac{b}{c} \right)^* \leq 
\left( \frac{ \gamma}{\gamma-1} \right) \kappa_0 g^{-1/(\gamma-1)}.
\end{equation}

Note that for all $\gamma>1$, as $g \to 1$, both bounds approach the harmonic mean Simpson degree, which is  $\kH=\kappa_0 \gamma/(\gamma-1)$.

\subsection{Sharper bounds for $g>1/2$}
\label{sec:sharperg}

\begin{figure}
\begin{center}
\begin{tabular}{cc}
\includegraphics[scale=0.7]{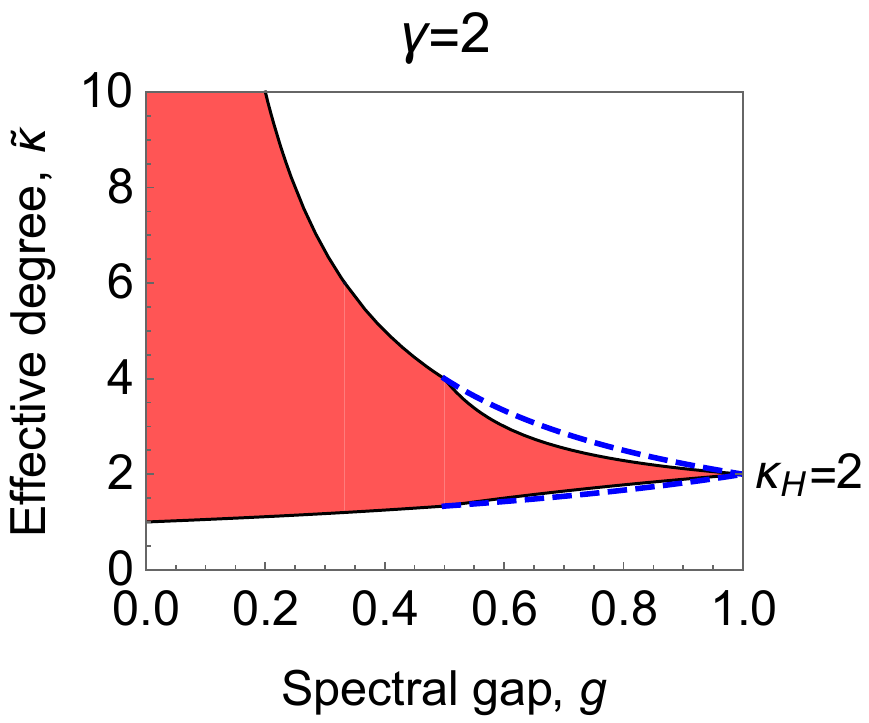} & \includegraphics[scale=0.7]{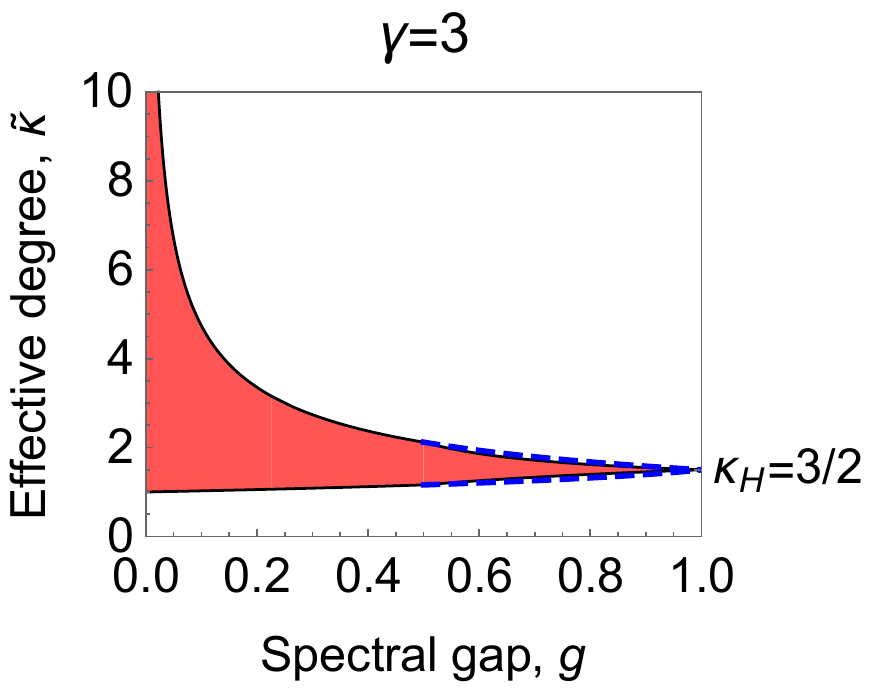}\\
\textsf{\textbf{A}} & \textsf{\textbf{B}}
\end{tabular}
\caption{Upper and lower bounds on the effective degree for graphs with power-law distribution of Simpson degrees: $f(\kappa) \propto \kappa^{-\gamma}$, for $\kappa \in [1, \infty)$. For $g<1/2$ we use the bounds \eqref{eq:powerlawbounds}, while for $g>1/2$ we use the sharper bounds \eqref{eq:powerlawboundsg}, showing also the bounds \eqref{eq:powerlawbounds} for comparison.}
\label{fig:powerlawgbounds}
\end{center}
\end{figure}

In the case $g>1/2$, we can use the bounds \eqref{eq:largegbounds2}.  From Eqs.~\eqref{eq:gammalower} and \eqref{eq:gammaupper}, we have
\begin{align*}
\klow & = \frac{\gamma \kappa_0}{2(\gamma-1) \big(1- 2^{-\gamma/(\gamma-1)}\big)},\\
\khigh & = \frac{\gamma \kappa_0}{2(\gamma-1) 2^{-\gamma/(\gamma-1)}}.
\end{align*}
This leads to
\begin{equation}
\label{eq:powerlawboundsg}
\left( \frac{\gamma}{\gamma-1} \right) \frac{g \kappa_0} {2g-1 + (1-g)2^{-1/(\gamma-1)}} 
\leq \left( \frac{b}{c} \right)^* \leq 
\left( \frac{\gamma}{\gamma-1} \right)
\frac{g \kappa_0}{ 1 - (1-g)2^{-1/(\gamma-1)}}.
\end{equation}
Figure \ref{fig:powerlawgbounds} illustrates these bounds for $\gamma=2$ and $\gamma=3$, both with $\kappa_0=1$.  In both cases, the improvement over bounds \eqref{eq:powerlawbounds} is relatively small.  

\section{Island model}
\label{app:island}

Here we describe an island model in which multiple isothermal graphs are joined together by weak links (Fig.~1D of the main text).  We begin with $n$ separate isothermal graphs (``islands") $G_1,\ldots,G_n$, of respective sizes $N_1,\ldots,N_n$. Each individual island is vertex-transitive, but the size and graph structure may vary across islands. Since each island is vertex-transitive, all vertices of a given island $G_x$ have the same Simpson degree $\kappa_x$ (but the $\kappa_x$ may differ of across islands). 

The islands are joined into an overall isothermal graph $G$ by adding an edge of weight $\alpha \ll 1$ between each pair of vertices on different islands.  To maintain a weighted degree of one, the edges within each island $G_x$ are rescaled by the factor $1-\alpha(N-N_x)$.

Since each island is vertex-transitive, and each inter-island pair is equally connected (by weight $\alpha$), the coalescence time $\tau_{ij}$ from different islands $G_x$ and $G_y$ depends only on $x$ and $y$, and not on the particular vertices $i \in G_x$ and $j \in G_y$.  Accordingly, we let $\tau_{G_xG_y}$ denote the meeting time between a vertex of $G_x$ and a vertex of $G_y$, $y \neq x$.  

From Eq.~\eqref{eq:taurecur} we have the following recurrence for coalescence times. For a pair $i \neq j$ on a common island $G_x$, we have
\begin{multline}
\label{eq:tauintra}
 \tau_{ij} = 1 + \frac{1}{2}\left(1-\alpha(N-N_x) \right) \sum_{k \in G_x} (w_{ik}\tau_{jk} + w_{jk}\tau_{ik})  
 +\sum_{y \neq x} \alpha N_y \tau_{G_xG_y}.
\end{multline}
For inter-island meeting times, we have that for $x \neq y$,
\begin{multline}
\label{eq:tauinter}
 \tau_{G_xG_y} = 1 + \frac{\alpha}{2 N_x} \sum_{i,j \in G_y} \tau_{ij} + \frac{\alpha }{2 N_y} \sum_{i,j \in G_x} \tau_{ij}\\
  + \frac{\alpha}{2} \sum_{ z \notin\{x,y\}} N_z \left( \tau_{G_z G_y}  +  \tau_{G_x G_z} \right) \\
+  \left(1-\frac{(N-N_x)\alpha}{2}-\frac{(N-N_y)\alpha}{2} \right) \tau_{G_x G_y}.
\end{multline}
We seek an asymptotic solution as $\alpha \to 0$.  As an \emph{ansatz}, we suppose
\begin{equation}
\begin{cases}
\tau_{ij} = \mathcal{O}(1) & \text{for $i \neq j$ and $i,j \in G_x$ for some $x$, }\\
\tau_{G_xG_y} = T_{xy}\alpha^{-1} + \mathcal{O}(1) & \text{for $x \neq y$},
\end{cases}
\end{equation}
for some collection of values $T_{xy}$. Substituting in Eq.~\eqref{eq:tauinter} and taking $\alpha \to 0$, we find that the $T_{xy}$ satisfy 
\begin{equation}
\label{eq:Tsystem1}
\frac{2N-N_x-N_y}{2} T_{xy} = 1 + \frac{1}{2}\sum_{ z \notin\{x,y\}} N_z(T_{xz} + T_{yz}).
\end{equation}
Defining $T_{xx}=0$ for all $x=1, \ldots, n$, we can rewrite Eq.~\eqref{eq:Tsystem1} as
\begin{equation}
\label{eq:Tsystem}
N T_{xy} = 1 + \frac{1}{2}\sum_{z=1}^n N_z(T_{xz} + T_{yz}).
\end{equation}
We have not found a general closed-form solution to Eq.~\eqref{eq:Tsystem}.  However, if there are $n=2$ islands, or if all islands have the same size $N/n$, the solution is $T_{xy}=n/N$ for all pairs $x,y$ with $x \neq y$.  

Taking $\alpha \to 0$ in Eq.~\eqref{eq:tauintra} yields
\begin{equation}
\tau_{ij} = 1 + \sum_{y=1}^n  N_y T_{xy} + \frac{1}{2} \sum_{k \in G_x} (w_{ik}\tau_{jk} + w_{jk}\tau_{ik}).
\end{equation}
This is the same as the recurrence \eqref{eq:taurecur} for coalescence times on island $G_x$ alone, except the time increment is 
$1 +\sum_{y \neq x}  N_y T_{xy}$ instead of 1. It follows that, when the islands are joined to form the overall graph $G$, all coalescence times from pairs $i,j \in G_x$ are scaled by the factor $1 + \sum_{y=1}^n  N_y T_{xy}$.  This scaling applies to remeeting times as well, so that if $\hat{\tau}_i$ is the remeeting time from vertex $i$ on graph $G_x$ alone, the remeeting time from $i$ on the overall graph $G$ is 
\[
\tau_i = \left( 1 + \sum_{y=1}^n N_y T_{xy} \right) \hat{\tau}_i.
\]

We check this solution by computing the overall sum of remeeting times, which should equal $N^2$ by Eq.~\eqref{eq:Kac}:
\begin{align}
\nonumber
\sum_{i \in G} \tau_i & = \sum_{x=1}^n \left( 1 + \sum_{y=1}^n  N_y T_{xy} \right) \sum_{i \in G_x}  \hat{\tau}_i\\
\nonumber
& = \sum_{x=1}^n \left( 1 + \sum_{y=1}^n  N_y T_{xy} \right)N_x^2 \\
\label{eq:taucheck}
& = \sum_{x=1}^n N_x^2  + \sum_{x, y} N_x^2 N_y T_{xy} .
\end{align}
To show this is equal to $N^2$, we multiply Eq.~\eqref{eq:Tsystem} by $N_xN_y$ and sum over all pairs $x,y$ with $x \neq y$:
\begin{align*}
N \sum_{\substack{x,y\\x \neq y}} N_x N_y T_{xy} & = \sum_{\substack{x,y\\x \neq y}} N_x N_y + \frac{1}{2}\sum_{\substack{x,y,z\\x \neq y}} N_x N_y N_z(T_{xz} + T_{yz})\\
N \sum_{x,y} N_x N_y T_{xy}
& = N-\sum_{x=1}^n N_x^2 + \frac{1}{2} 
\sum_{x,y,z} N_x N_y N_z(T_{xz} + T_{yz}) - \sum_{x,z} N_x^2 N_z T_{xz}\\
& = N-\sum_{x=1}^n N_x^2 + N \sum_{x,y} N_x N_y T_{xy} - \sum_{x,y} N_x^2 N_y T_{xy}.
\end{align*}
This yields the identity
\begin{equation}
\label{eq:NTsum}
\sum_{x,y} N_x^2 N_y T_{xy}= N^2 - \sum_{x=1}^n N_x^2.
\end{equation}
Substituting in Eq.~\eqref{eq:taucheck} yields $\sum_{i \in G} \tau_i = N^2$ as desired.

We compute the effective degree $\tilde{\kappa}$ of $G$ as
\begin{align*}
\tilde{\kappa} 
& = N^2 \left(\sum_{x=1}^n \sum_{i \in G_x} \tau_i \kappa_i^{-1} \right)^{-1}\\
& = N^2 \left(\sum_{x=1}^n \left( 1 + \sum_{y=1}^n N_y T_{xy} \right) \kappa_x^{-1} \sum_{i \in G_x} \hat{\tau}_i  \right)^{-1}\\
& = N^2 \left(\sum_{x=1}^n \left( 1 + \sum_{y=1}^n N_y T_{xy} \right) N_x^2 \kappa_x^{-1}  \right)^{-1}.
\end{align*}
So the effective degree $\tilde{\kappa}$ is a weighted harmonic average of the Simpson degrees on the separate islands, with each $\kappa_x$ weighted by $ N_x^2\left( 1 + \sum_{y=1}^n N_y T_{xy} \right)$.  

In the case that all islands have equal size, ($N_x=N/n$ for all $x$), the islands are weighted equally.  For $n=2$ islands (not necessarily of equal size), substituting the solution $T_{12} = 2/N$ to Eq.~\eqref{eq:Tsystem}, we obtain
\begin{align*}
\tilde{\kappa} & = N^2 \left( N_1^2 (1+2N_2/N) \kappa_1^{-1} + N_2^2 (1+2N_1/N) \kappa_2^{-1} \right)\\
& = N^3 \left( N_1^2 (N_1 + 3N_2) \kappa_1^{-1} +  N_2^2 (3N_1 + N_2) \kappa_2^{-1}  \right)^{-1}.
\end{align*}

\section{Wheel graph}
\label{app:wheel}

The isothermal wheel graph (Fig.~1B of the main text) has $n$ wheel vertices and one hub.  Neighboring wheel vertices are joined by edges of weight $(n-1)/(2n)$, and each wheel vertex is joined to the hub by an edge of weight $1/n$.  Let $\tau_{L,j}$ denote the coalescence time for two leaves that are $j$ apart, $0 \leq j \leq n$.  Clearly, $\tau_{L,0}  = \tau_{L,n} = 0$.  Let $\tau_{LH}$ denote the coalescence time between a leaf and the hub.

Recurrence relation \eqref{eq:taurecur} for coalescence times becomes
\begin{align}
\label{eq:tauLj}
\tau_{L,j} & = 1 + \frac{n-1}{2n} \left( \tau_{L,j-1} + \tau_{L,j+1} \right) + \frac{1}{n} \tau_{LH} \quad \text{for $1 \leq j \leq n-1$},\\
\label{eq:tauH}
\tau_{LH} & = 1 + \frac{n-1}{2n} \tau_{LH} + \frac{1}{2n} \sum_{j=0}^{n-1} \tau_{L,j}.
\end{align}
For convenience, we define $\tau_{L,j}'=\tau_{L,j}-\tau_{LH}$.  Then Eqs.~\eqref{eq:tauLj}--\eqref{eq:tauH} become
\begin{align}
\label{eq:tauLjprime}
\tau_{L,j}' & = 1 + \frac{n-1}{2n} \left( \tau_{L,j-1}' + \tau_{L,j+1}' \right)  \quad \text{for $1 \leq j \leq n-1$},\\
\label{eq:taujsum}
\tau_{LH} & = 2n + \sum_{j=0}^{n-1} \tau_{L,j}'.
\end{align}
As an \emph{ansatz}, we suppose the solution to Eq.~\eqref{eq:tauLjprime} takes the form
\begin{equation}
\label{eq:wheelansatz}
\tau_{L,j}' = a  + b\left( \gamma^j + \gamma^{n-j} \right),
\end{equation}
for some $a,b,\gamma$ depending on $n$ but not on $j$. Substituting into Eq.~\eqref{eq:tauLjprime} gives
\begin{align*}
a  + b\left( \gamma^j + \gamma^{n-j} \right) & = 
1 + \frac{n-1}{2n} \left( 2a + b \left( \gamma^{j-1} + \gamma^{n-j+1} + \gamma^{j+1} + \gamma^{n-j-1} \right) \right)\\
& = 1 + \frac{n-1}{2n} \left( 2a + b \left( \gamma + \gamma^{-1} \right) \left( \gamma^j + \gamma^{n-j} \right) \right).
\end{align*}
For this to hold for all $1 \leq j \leq n-1$ necessitates that
\[
a = 1 + \frac{n-1}{n}a \quad \text{and} \quad (n-1) \left(\gamma + \gamma^{-1} \right)= 2n.
\]
Solving the above equations yields $a=n$ and 
\[
\gamma = \frac{n - \sqrt{2n-1}}{n-1}, \qquad \gamma^{-1} = \frac{n + \sqrt{2n-1}}{n-1}.
\]
To solve for $b$, we substitute into Eq.~\eqref{eq:taujsum}, 
\begin{align}
\nonumber
\tau_{LH} & = 2n +  \sum_{j=0}^{n-1} \left(n + b\left( \gamma^j + \gamma^{n-j} \right) \right)\\
\label{eq:tauLHb}
& = n(n+2) + b \frac{(1+\gamma)(1-\gamma^n)}{1-\gamma}.
\end{align}
Additionally, since $\tau_{L,0}=0$, we have 
\begin{equation}
\label{eq:tauLHb2}
\tau_{LH}  = -\tau_{L,0}' = -n -b(1+\gamma^n).
\end{equation}
Combining Eqs.~\eqref{eq:tauLHb} and \eqref{eq:tauLHb2} and solving for $b$ yields
\[
b  = -\frac{n(n+3)}{2} \left(\frac{1-\gamma}{1-\gamma^{n+1}} \right).
\]

Substituting this value of $b$ into Eqs.~\eqref{eq:tauLHb2} and \eqref{eq:wheelansatz}, we obtain the coalescence times
\begin{align*}
\tau_{LH} & = \frac{n(n+3)}{2} \left(\frac{(1-\gamma)(1+\gamma^n)}{1-\gamma^{n+1}} \right) - n,\\
\tau_{L,j} & = \tau_{L,j}' +\tau_{LH} \\
& = \frac{n(n+3)}{2} \left(\frac{(1-\gamma)(1-\gamma^j)(1-\gamma^{n-j})}{1-\gamma^{n+1}} \right).
\end{align*}
In particular, for neighboring leaves ($j=1$), we have
\begin{equation}
\tau_{L,1} =  \frac{n(n+3)}{2} \left(\frac{(1-\gamma)^2(1-\gamma^{n-1})}{1-\gamma^{n+1}} \right).
\end{equation}

Turning now to remeeting times, we compute
\begin{align*}
\tau_H & = 1 + \tau_{LH}\\ 
& =  \frac{n(n+3)}{2} \left(\frac{(1-\gamma)(1+\gamma^n)}{1-\gamma^{n+1}} \right) - (n-1),\\
\tau_L & = 1 + \tfrac{1}{n}\tau_{LH} + \tfrac{n-1}{n} \tau_{L,1} \\
& =  \frac{n+3}{2} \left(\frac{(1-\gamma)(n(1+\gamma^n)-(n-1)(\gamma+\gamma^{n-1}))}{1-\gamma^{n+1}} \right).
\end{align*}
The Simpson degrees are
\[
\kappa_H = n, \qquad \kappa_L = \left( 2 \frac{(n-1)^2}{4n^2} + \frac{1}{n^2} \right)^{-1}= \frac{2n^2}{(n-1)^2+2}.
\]
Using the above values, the effective degree can be calculated as 
\begin{equation}
\label{eq:kwheel}
\tilde{\kappa}  = \frac{(n+1)^2}{ \tau_H\kappa_H^{-1}  + n \tau_L \kappa_L^{-1}}.
\end{equation}

Asymptotically, as $n \to \infty$, we have 
\[
\tau_H \sim \frac{n\sqrt{n}}{\sqrt{2}},  \qquad \kappa_H \sim n, \qquad \tau_L \sim n, \qquad \kappa_L \sim 2,
\]
where $f(n) \sim g(n)$ means that $\lim_{n \to \infty} f(n)/g(n)=1$. (The asymptotic expressions for $\tau_H$ and $\tau_L$ were obtained with the aid of Mathematica.) Substituting into Eq.~\eqref{eq:kwheel} and simplifying gives $\lim_{n \to \infty} \tilde{\kappa} = 2$.

\section{Numerical experiments for power-law networks}
\label{app:numerical}

We have computed the effective degree $\tilde{\kappa}$ and critical $B/C$ ratio for two families of random graphs on $N=200$ nodes: (a) power-law generated via the configuration model, (b) preferential attachment.  

For the configuration model, we first generated a degree sequence by rounding down a sequence of random reals chosen from the interval $[ k(\gamma-2)/(\gamma-1), \infty )$ according to a $\gamma$-exponent power-law distribution. Then we generated a random graph with the given degree sequence using the standard configuration model. 

For the preferential attachment graphs, we start from a complete graph of size $(m+1)$, and add vertices one at a time. Each new vertex connects to exactly $m$ distinct previous vertices, with probability proportional to $(\mathrm{degree} + m(\gamma-3))$. This procedure grows a power-law network with exponent $\gamma$ and mean degree $k = 2m$ \cite{dorogovtsev2000structure,krapivsky2001organization}.

For each graph topology we generated, we found an isothermal weighting consistent with this topology by numerically solving a quadratic program to minimize $\sum_{i,j} w_{ij}^2$ under the constraint $\sum_j w_{ij}=1$ for all $i$. (``Consistent with this topology" means that only the edges that are present in the given topology are allowed to have nonzero weights.) The sum-of-squares minimization achieves a weighting as uniform as possible, and avoids biasing the results in favor of cooperation.

In both cases we let the asymptotic average degree, $k=2m$, range from 6 to 24 in increments of 2, and the power-law exponent $\gamma$  from $2.05$ to $3$ in increments of 0.05. For each parameter combination, 10 random graphs were generated, for a total of 2000 graphs in each ensemble.

For some topologies, an isothermal weighting may not exist.  For instance, if a set $L$ of low-degree nodes happen to be only connected to the same subset $H$ of hubs, and $|L| > |H|$, then it is clear that there cannot be an isothermal weighting. Such configurations arise with non-negligible probability for small values of $\gamma$. We removed such graphs from the ensemble.  Consequently, the resulting sample does not have the same number of graphs in each category.

\end{document}